\pdfoutput=1 
\documentclass[aps,prl,twocolumn,groupedaddress,amsmath]{revtex4}

\usepackage{bm} 

\usepackage{comment}
\usepackage{graphicx}
\usepackage{amssymb}

\usepackage{times} 
\usepackage{amsmath,amsfonts,latexsym}

\newcommand*{\cU}{{\cal U}}

\newcommand*{\cJ}{{\cal J}}

\newcommand*{\Pop}{\Psi^{\vphantom{\dagger}}}
\newcommand*{\Pdop}{\Psi^\dagger}

\newcommand*{\phop}{\phi^{\vphantom{\dagger}}}

\newcommand*{\sgn}{\mathrm{sgn}}

\begin{document}

\title{Dispersion relation and spectral function of an impurity in a one-dimensional quantum liquid}

\author{Austen Lamacraft} 
\email{austen@virginia.edu}
\affiliation{Department of Physics, University of Virginia,
Charlottesville, VA 22904-4714 USA}
 
\date{\today}

\begin{abstract}
We consider the motion of an impurity particle in a general one-dimensional quantum fluid at zero temperature. The dispersion relation $\Omega(P)$ of the impurity is strongly affected by interactions with the fluid as the momentum approaches $\pm\pi\hbar n, \pm 3\pi\hbar n, \ldots$, where $n$ is the density. This behavior is caused by singular $\pm 2\pi\hbar n$ scattering processes and can be understood by analogy to the Kondo effect, both at strong and weak coupling, with the possibility of a quantum phase transition where $\Omega'(\pm \pi n)$ jumps to zero with increasing coupling. The low energy singularities in the impurity spectral function can be understood on the same footing.
\end{abstract}

\maketitle

The study of systems in which one or a few distinguished degrees of freedom are simultaneously coupled to a `bath' consisting of an infinite number more finds applications in all branches of physics. In condensed matter physics, such situations are often called `impurity' problems, since the canonical examples describe the dynamics of a defect in an otherwise perfect crystal lattice, or a foreign atom in a pure fluid. 

A natural distinction within the class of impurity problems is between the situation typical of the solid state, in which impurities are fixed and have infinite mass, and that found more often in the study of quantum liquids such as superfluid $^{4}$He, where the impurity is mobile and has finite mass. While the former has historically provided a richer source of new concepts, including the related phenomena of the Kondo effect~\cite{kondo1964}, the X-ray edge problem~\cite{mahan1967,nozieres1969} and Anderson's orthogonality catastrophe~\cite{anderson1967}, the latter finds new applications in the physics of ultracold atomic gases. Ref.~\cite{chikkatur2000} provides an example of this new setting. A small fraction of atoms in a Bose condensed gas were transferred into a different hyperfine state, forming a dilute system of impurities propagating through the condensate. With the momentum and energy of the impurities under independent control, a more sophisticated version of the same experiment could measure the spectral functions discussed later in this work.

Recent experimental advances mean that the study of mobile impurities in low dimensional, strongly interacting atomic gases has become a possibility~\cite{bloch2008}. The purpose of this paper is to show that in one dimension such a system displays remarkably rich behavior. The model that we will discuss applies equally well to solid state systems such as quantum wires and carbon nanotubes in which the gas corresponds to a partially filled sub-band and the impurities to carriers propagating in an unfilled sub-band.
\begin{figure}
\centering  \includegraphics[width=0.5\textwidth]{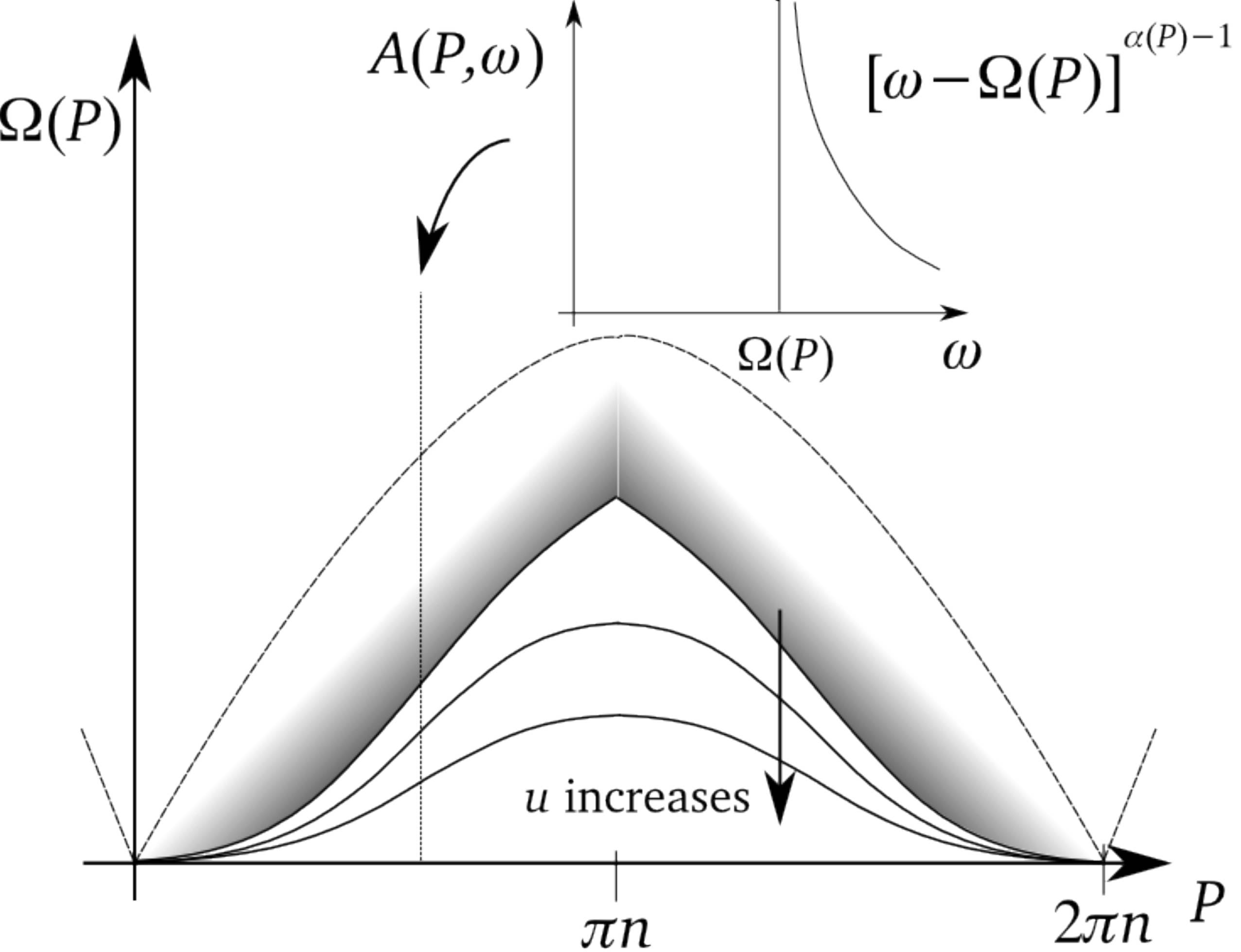}
\caption{Schematic dispersion relation of an impurity moving in $K>1$ Luttinger liquid. At weak coupling a cusp persists at $P=\pi n$, but vanishes discontinuously at some critical coupling. In the strong coupling limit the dispersion takes the simple form $\Omega(P)\propto\sin^2 P/2n$. $\Omega(P)$ is the ground state energy of the system at given $P$: the singular form of the impurity spectral function near this threshold is indicated in the inset. The dashed curve illustrates the ground state in the absence of an impurity for reference.
 \label{fig:schem}}
\end{figure}

The problem to be considered can be motivated by the following simple calculation. Consider an impurity of mass $M$ moving through a Fermi gas consisting of particles of mass $m$ and density $n$. With an eye to applications in ultracold physics the interaction between the gas and the impurity will be taken to be $H_{\mathrm{int}}=u\sum_i\delta(x_i-X)$. We take $u>0$: for attractive interactions a bound state will form and we expect the problem is similar but with a `molecular' impurity. Let us find how the dispersion relation $\Omega^{(0)}(P)=P^2/2M$ of the impurity is modified by this interaction. At first order in $u$ the contribution $\Omega^{(1)}=un$ is momentum independent, while the second order contribution is (we set $\hbar=1$)
\begin{eqnarray*}
\Omega^{(2)}(P)&=&-\frac{u^2}{L^2}\mathop{\sum_{|p_1|>p_F}}_{|p_2|<p_F}\frac{1}{\frac{\left(P-p_1+p_2\right)^2}{2M}+\frac{p_1^2}{2m}-\frac{p_2^2}{2m}-\frac{P^2}{2M}}\\\nonumber
\end{eqnarray*}
%
Near the Fermi points $\pm p_F=\pm \pi n$ we have the singular behavior
\begin{equation}\label{sing}
\Omega^{(2)}(P)\sim \mathrm{const.}+\frac{u^2m^2M|p_F-|P||}{2\pi^2(M^2-m^2)}\ln\left|\frac{p_F}{p_F-|P|}\right|+\cdots,
\end{equation}
so that the derivative $\Omega'(P)$ has a logarithmic singularity at the Fermi points. The case $M=m$ is special, and we have
\begin{equation}\label{equal_PT}
\Omega^{(2)}(P)=-\frac{mu^2}{8\pi^2}\left[\ln^2\left(\frac{p_F+P}{p_F-P}\right)+\pi^2\right],\, |P|<p_F
\end{equation}
In both cases the contribution is negative. The origin of the singular behavior in Eq.~(\ref{sing}) as $P\to \pm p_F$ is the following. There are intermediate states in the second order calculation involving a soft particle hole pair with momentum $\sim 0$, leading to a vanishing denominator. Additionally, when $P\to \pm p_F$, a vanishing denominator arises for an intermediate state in which the impurity is scattered to the opposite Fermi point, accompanied by a particle hole pair of momentum $\sim \pm 2p_F$.  
%
%
We will see that this is analogous to the contribution of spin-flip processes at second order in the Kondo problem and corresponds to identical singular behavior as a function of magnetic field. 

In the above calculation, the meaning of $\Omega(P)$ is clear, being the energy of a state that goes over  to the free impurity with momentum $P$ at zero coupling. How should we understand $\Omega(P)$ in the general case, without reference to perturbation theory? A kinematic argument shows that the ground state energy of the system with total momentum $P$ may be identified with the impurity dispersion as long as the group velocity $|\Omega'(P)|<v$, the speed of sound (equal to the Fermi velocity $v_F$ in this case), 
otherwise a lower energy state will be one with real particle-hole or phonon excitations. Since ${\Omega^{(0)}}'(p_F)=mv_F/M$ the above calculation is then meaningful for $M\geq m$. By defining $\Omega(P)$ as the ground state energy at given $P$ we obtain a periodic function $\Omega(P+2\pi n)=\Omega(P)$. 


The remainder of this paper is concerned with a general 1D quantum fluid characterized by a Luttinger parameter $K$, equal to  $\pi n/m v$ for a Galilean invariant system~\cite{Haldane1981}.  The main findings for $\Omega(P)$ are then (see Fig.~\ref{fig:schem}):

\begin{enumerate}
 
\item For $K\leq 1$, including the case of free fermions with $K=1$, $\Omega'(\pm \pi n)=0$ for \emph{arbitrarily weak} coupling between the impurity and gas.

\item For $K>1$ (e.g. a Bose gas), $\Omega'(\pm \pi n)$ is finite for weak coupling, but jumps discontinuously to zero at some critical coupling.

\item At strong coupling, when the impurity is almost opaque to the gas particles, the dispersion takes the simple form $\Omega(P)-\Omega(0)\propto\sin^2 P/2n$, with the bandwidth being set by the  amplitude for tunneling through the impurity.

\end{enumerate}

Note that the impurity propagates ballistically only at zero temperature, with a finite mobility diverging as $\propto T^{-4}$ at low temperatures~\cite{kagan1986epe,castro-neto1994,castroneto1996}.

In addition we will find the form of the spectral function $A^+(P,\omega)$ for the addition of the impurity at momentum $P$.
Since $\Omega(P)$ is the ground state of the system with total momentum $P$, this energy represents a threshold below which the spectral function vanishes. Just above the threshold the spectral function has the form of a power law with momentum-dependent exponent
\begin{equation}\label{gen_spect}
A^+(P,\omega)\propto \theta(\omega-\Omega(P))\left[\omega-\Omega(P)\right]^{\alpha(P)-1}.
\end{equation}
In general the calculation of the exponent $\alpha(P)$ is a difficult task. In the limit of an almost opaque impurity we obtain $\alpha(P)=\frac{K}{2}[\left(\frac{U_\phi}{\pi v}\right)^2+\left(\frac{P}{\pi n}\right)^2]$, where $U_\phi$ is a non-universal parameter, equal to $\pi v$ in the Fermi or Tonks gas. Finally, we confirm those results that apply to the $K=1$ case for the Fermi gas with $M=m$, which is a particularly simple integrable system.


We begin be developing the theory at weak coupling. As indicated above, it is convenient to describe the low energy degrees of freedom of the gas as a Luttinger liquid with Hamiltonian~\cite{Haldane1981}
\begin{equation}\label{luttinger}
H_{\mathrm{gas}}=\frac{v}{2\pi}\int dx\left[K\left(\partial_x\theta\right)^2+\frac{1}{K}\left(\partial_x\phi\right)^2\right]
\end{equation}
with $\left[\phi(x),\partial_y\theta(y)\right]=i\pi\delta(x-y)$. The total Hamiltonian of the system is then $H_{\mathrm{imp}}+H_{\mathrm{gas}}+u\rho(X)$, where  $H_{\mathrm{imp}}=P^2/2M$, $\left[P,X\right]=-i$ and $\rho(x)$ is the density of the gas. In the Luttinger liquid picture this has the form $\rho(x)=\frac{1}{\pi}\partial_x\phi(x)\sum_{m=-\infty}^\infty e^{2mi\phi(x)}$.

It is convenient to make a transformation to a frame moving with the impurity, which can be accomplished by the unitary transformation $\cU_X=e^{iXP_{\mathrm{gas}}}$, where $P_{\mathrm{gas}}$ is the momentum of the gas. Then $H=H_{\mathrm{gas}}+(P-P_{\mathrm{gas}})^2/2M+u\rho(0)$. With $X$ now absent from the Hamiltonian $P$ is conserved and corresponds to the total momentum of the system. Neglecting irrelevant operators arising from higher order harmonics of the density, the low energy Hamiltonian then takes the form
%
\begin{eqnarray}\label{luttinger_impurity2}
H&=&\frac{v}{2\pi}\int dx\left[K\left(\partial_x\tilde\theta\right)^2+\frac{1}{K}\left(\partial_x\tilde\phi\right)^2\right]\nonumber\\
&&+\frac{1}{2M}\left(P-\pi n J-\frac{1}{\pi}\int dx\, \partial_x\tilde\phi\partial_x\tilde\theta\right)^2\nonumber\\
&&+\frac{U_\phi}{\pi}\partial_x\tilde\phop(0)+2nU_{2\pi n}\cos\left[2\left(\phop_0+\tilde\phop(0)\right)\right].
\end{eqnarray}
%
In writing Eq.~(\ref{luttinger_impurity2}) we have passed from the microscopic interaction strength $u$ to effective couplings $U_\phi$ and $U_{2\pi n}$, describing forward scattering and backward scattering from the impurity respectively. For the $\delta$-function interaction we have  $U_\phi=U_{2\pi n}=u$ at lowest order in $u$.
We have also separated the zero mode contributions in $\phi(x)$ and $\theta(x)$.
\begin{eqnarray}\label{zero_split}
\phi(x)&=&\phi_0+\frac{\pi Nx}{L}+\tilde\phi(x)\nonumber\\
\theta(x)&=&\theta_0+\frac{\pi Jx}{L}+\tilde\theta(x).
\end{eqnarray}
with $\left[\phi_0,J\right]=\left[\theta_0,N\right]=i$. The backward scattering term thus changes $J$ by 2, corresponding to a $\pm 2\pi n$ change in the momentum of the gas. If $\pi n\left[\cJ-1\right]<P<\pi n\left[\cJ+1\right]$ for even integer $\cJ$, the second term of Eq.~(\ref{luttinger_impurity2}), arising from the kinetic energy of the impurity, favors the value $J=\cJ$. For $P=\pi n\left(\cJ+1\right)$, states corresponding to $J=\cJ,\cJ+2$ become degenerate in the absence of backward scattering; away from this point they have a splitting $\Delta_{P,\cJ}\equiv\frac{2\pi n}{M}\left[P-\pi n\left(\cJ+1\right)\right]$.

Before considering the structure of perturbation theory in $U_{2\pi n}$ it is convenient to first remove the forward scattering terms by a unitary transformation  $\cU_L\cU_R$ with~\cite{gogolin1998}
\begin{eqnarray}\label{luttinger_trans}
\cU_{L/R}=\exp\left[\pm i\frac{\sqrt{K}U_\phi\tilde\phop_{L/R}(0)}{2\pi\left(v\mp P_J/M\right)}\right]
\end{eqnarray}
where $P_J=P-\pi n J$ and we introduced the chiral modes
%
$\tilde\phop_{L/R}=\tilde\phi/\sqrt{K}\pm\sqrt{K}\tilde\theta$,
%
%
obeying the relations $[\tilde\phop_{L/R}(x),\partial_{x'}\tilde\phop_{L/R}(x')]=\pm 2\pi i \delta_{ss'}\delta(x-x')$. 
%
%
Close to a degeneracy point we retain only the values $J=\cJ, \cJ+2$, so that Hamiltonian is conveniently written using the pseudo-spin 1/2 variable $\sigma^z=J-\cJ-1$ so that $P_J=-\pi n \sigma^z$ and
\begin{widetext}
\begin{eqnarray}\label{luttinger_kondo}
H=\frac{1}{4\pi}\int dx\left[(v-\pi n \sigma^z/M )(\partial_x\tilde\phop_R)^2+(v+\pi n \sigma^z/M )(\partial_x\tilde\phop_L)^2\right]
-\frac{\tilde\Delta_{P,\cJ}}{2}\sigma^z+2nU_{2\pi n}\left[\sigma_+e^{i\left(\eta\left[\tilde\phop_L(0)+\tilde\phop_R(0)\right]\right)}+\mathrm{h.c.}\right]
\end{eqnarray}
\end{widetext}
where
\begin{equation}\label{etas}
\eta=\sqrt{K}\left[1-\frac{nMU_\phi}{M^2v^2-\pi^2n^2}\right],
\end{equation}
%
and $\tilde\Delta_{P,\cJ}$ indicates that the splitting is altered from its bare value in a way that depends upon the short distance behavior of the problem and not specified by the Luttinger Hamiltonian. The precise value is not important in what follows. We have also dropped the term $\frac{1}{2M\pi^2}\left(\int dx\, \partial_x\tilde\phi\partial_x\tilde\theta\right)^2$ in Eq.~(\ref{luttinger_impurity2}) as it gives is an irrelevant operator that does not change our conclusions.

$\Omega(P)$ for $P\sim\pi n(\cJ+1)$ is just the ground state energy of Eq.~(\ref{luttinger_kondo}) for $\tilde\Delta_{P,\cJ}\sim 0$. In particular, the presence of a cusp in $\Omega(P)$ corresponds to a jump in the expectation value $\langle \sigma^z\rangle\propto \Omega'(P)$ as $P$ passes $\pi n(\cJ+1)$, as occurs for $U_{2\pi n}=0$. Let us discuss the behavior of the system as a function of $U_{2\pi n}$ and $\eta$, before returning to our original parameters. 

A cusp is preserved for small $U_{2\pi n}$ if the backward scattering term is irrelevant, i.e. for $\eta^2>1$. To understand what happens for larger $U_{2\pi n}$, or for $\eta^2<1$, we observe that an expansion of the ground state energy of Eq.~(\ref{luttinger_kondo}) $U_{2\pi n}$ yields the Anderson-Yuval expansion~\cite{Yuval1970} of the Kondo problem in a magnetic field proportional to $\tilde\Delta_{P,\cJ}$~\footnote{The `Doppler shifts' of the mode velocities that appear in the first line of Eq.~(\ref{luttinger_kondo}) only serve to alter the coupling by a factor $\left[1-\left(\pi n/Mv\right)^2\right]^{-\eta^2/2}$}. For $\eta^2<1$ backward scattering is relevant and the ground state corresponds to the antiferromagnetic Kondo problem i.e. a singlet with $\langle \sigma^z\rangle=0$ and a finite impurity susceptibility, corresponding to $\Omega'(\pi n(\cJ+1))=0$, $\Omega''(\pi n(\cJ+1))<0$. For $\eta^2>1$, $\langle \sigma^z\rangle$ jumps discontinously to zero from the universal value $\eta^{-1}$ as $u$ increases past some critical value~\cite{Imbrie1988} and the Kondo ground state switches from ferromagnetic to antiferromagnetic.


In terms of our original parameters, this means that for $K<1$,  $\Omega'(\pi n(\cJ+1))=0$ always, while for $K>1$ a transition occurs with  $\Omega'(\pi n(\cJ+1))$ jumping to zero as $u$ is increased. The determination of the critical value of $u$ is a difficult problem that we do not address here. The divergence of Eq.~(\ref{etas})  as $M\to \pi n/v$ is to be expected, as in this limit the bare group velocity at momentum $\pm \pi n$ equals the sound velocity. This limitation does not exist for the strong coupling limit where the dispersion is almost flat, as we will now show.

As $u\to\infty$ the impurity presents a hard wall to the particles of the gas. If the impurity's mass were infinite, this would correspond to the boundary conditions
%
$\phi(0_\pm)=\mp\frac{KU_\phi}{2v}$
%
The equations of motion $\dot\theta=v\partial_x\phi$, $\dot\phi=v\partial_x\theta$ then show that $\partial_x\theta$ vanishes as the origin is approached from either side, corresponding to vanishing current, while $\theta(x)$ is discontinuous at the origin: the phase difference can fluctuate wildly as the gas is cut in two. The value to which $\phi$ is pinned is determined by the magnitude of the forward scattering terms in Eq.~(\ref{luttinger_impurity2}) in the $u\to\infty$ limit and is \emph{non-universal}. In the simple case of a free Fermi gas or Tonks gas ($K=1$) we have $U_\phi=v\pi$.

In order to examine the effect of the impurity kinetic energy, we write the total momentum of the gas as
\begin{equation*}\label{momentum}
P_\mathrm{gas}=\frac{1}{\pi}\int_{|x|>\epsilon} dx\, \partial_x\theta\partial_x\phi=-n\theta(x)|^{0_+}_{0_-}+\frac{1}{\pi}\int_{|x|>\epsilon}dx\,\partial_x\theta\partial_x\tilde\phi
\end{equation*}
The origin is excluded as there is a break in the fluid here. The second term will be unimportant at low energies, so that we can take the impurity term in the Hamiltonian to be
\[H_{\mathrm{imp}}=\frac{1}{2M}\left(P+n\theta(x)|^{0_+}_{0_-}\right)^2.\]
The overall Hamiltonian $H_{\mathrm{gas}}+H_{\mathrm{imp}}$ is quadratic, but without discussing the solution explicitly we note $\langle\theta(x)|^{0_+}_{0_-}\rangle_{H_{\mathrm{gas}}+H_{\mathrm{imp}}}=-P/n$ with fluctuations about this value vanishing at energies $\lesssim n^2/M$ due to the pinning effect of $H_{\mathrm{imp}}$. Note that in this limit the ground state energy has no dependence on $P$, i.e. the impurity dispersion is flat.

Turning now to the case of small but finite transparency, the term in Hamiltonian allowing tunneling of particles of the gas through the impurity is the Josephson-like term $H_t=-2t\cos[\theta(0_+)-\theta(0_-)]$. 
The first order correction to the energy is then just the expectation value $\langle H_t\rangle=-2t\langle\cos[\theta(0_+)-\theta(0_-)]\rangle\sim-2t\left(m/M\right)^{1/K}\cos P/n$, so that the dispersion has the form advertised earlier.


We next discuss the behavior of the spectral function. The spectral function $A^+(P,\omega)$ is the Fourier transform of $\langle 0| \Pop_P(t)\Pdop_P(0)|0\rangle$, where the operator $\Pdop_P$ creates an impurity with momentum $P$, and $|0\rangle$ denotes the $P=0$ ground state of the gas in the absence of the impurity. Let us first consider the strong coupling case. We have established that the impurity kinetic energy $H_{\mathrm{imp}}$ ensures that at energies $\lesssim n^2/M$ the system is described by the simple boundary conditions
\begin{eqnarray}\label{low_bc}
\theta|^{0_+}_{0_-}&=&-P/n\nonumber\\
\phi|^{0_+}_{0_-}&=&-\frac{KU_\phi}{v}.
\end{eqnarray}
%
These boundary condition can be imposed on fields that are continuous at the origin by the unitary transformation 
\begin{equation}\label{UP}
\cU_{P}=\exp\left(\frac{iP}{\pi n}\phop(0)+\frac{iKU_\phi}{\pi v}\theta(0)\right),
\end{equation}
so that we have 
\[\langle 0| \Pop_P(t)\Pdop_P(0)|0\rangle=\theta(t)e^{-i\Omega(P)t}\langle \cU^\dagger_{P}(t)\cU^{\vphantom{\dagger}}_{P}(0)\rangle.\]
Since $\langle \cU^\dagger_{P}(t)\cU^{\vphantom{\dagger}}_{P}(0)\rangle\propto |t|^{-\frac{K}{2}[\left(\frac{U_\phi}{\pi v}\right)^2+\left(\frac{P}{\pi n}\right)^2]}$, the spectral function has the form Eq.~(\ref{gen_spect}) with $\alpha(P)$ as given earlier. The result agrees with Ref.~\cite{Matveev2008}, which treated the special case $M=m$, $K=1$ (see also Ref.~\cite{Zvonarev2007}).

Finally, we illustrate this behavior for the case of free fermions coupled to an impurity with $M=m$. This system is integrable, being an extreme limit of the spin-1/2 Fermi gas~\cite{McGuire2004a,Gaudin1967,Yang1967}. As shown in Ref.~\cite{castella1993}, an eigenstate of a system of $N$ fermions can be written in the co-moving frame as a Slater determinant of functions $\varphi_j(x)$ $j=1,\cdots N$ satisfying $\varphi_j(0)=\varphi_j(L)$ and $\varphi'_j(0)-\varphi_j'(L)=mu
\varphi_j(0)$ (the latter involving the reduced mass $m/2$, different from a static scatterer). The $\varphi_j(x)$ are expanded in terms of the \emph{$N+1$} plane wave states: $\varphi_j(x)=
\sum_{t=0}^Na_j^te^{ik_t x}$, where the momenta satisfy $k_j L=2\pi n_j-2\delta(k_j)$ with integers  $n_j$ and phase shift
\[\delta(k)=-\frac{\pi}{2}\sgn(k)+\arctan\left(\frac{2(k-\Lambda)}{mu}\right).\]
The lab frame energy  and momentum are $E=\frac{1}{2m}\sum_{j=0}^N k_j^2$, the $P=\sum_{j=0}^N k_j$, so that the spectral parameter $\Lambda$ gives the former as an implicit function of the latter. In the  limit $L\to\infty$ we find $E(\Lambda)=p_F^3L/(6\pi m)+\Omega(\Lambda)$ with
\begin{eqnarray}\label{BA_CZ}
\Omega(\Lambda)&=&\frac{p_F^2}{2m}+\int_{-p_F}^{p_F}\frac{dk}{2\pi}\frac{4mu}{(mu)^2+4(k-\Lambda)^2}\left[\frac{k^2}{2m}-\frac{p_F^2}{2m}\right]\nonumber \\
P(\Lambda)&=&-2\int_{-p_F}^{p_F}\frac{dk}{2\pi} \arctan\left(\frac{2(k-\Lambda)}{mu}\right)
\end{eqnarray}
We find the following i) Eq.~(\ref{equal_PT}) holds for $u\to 0$, ii) $\Omega(P)=\frac{p_F^2}{2m}-\frac{4p_F^3}{3\pi m^2 u}\cos^2\left(\pi P/2p_F\right)$  for $u\to \infty$ or $p_F\to 0$, and iii) $\Omega(P)\to \frac{p_F^2}{2m}-\frac{1}{3}\frac{p_F\pi(p_F-|P|)^2}{m^2u}$ as $P\to \pm p_F$. Thus $\Omega'(\pm p_F)=0$ for arbitrary $u$. The fact that $\Omega(\pm p_F)=p_F^2/2m$ for all $u$ is a consequence of $SU(2)$ symmetry.

\begin{figure}
\centering  \includegraphics[width=0.4\textwidth]{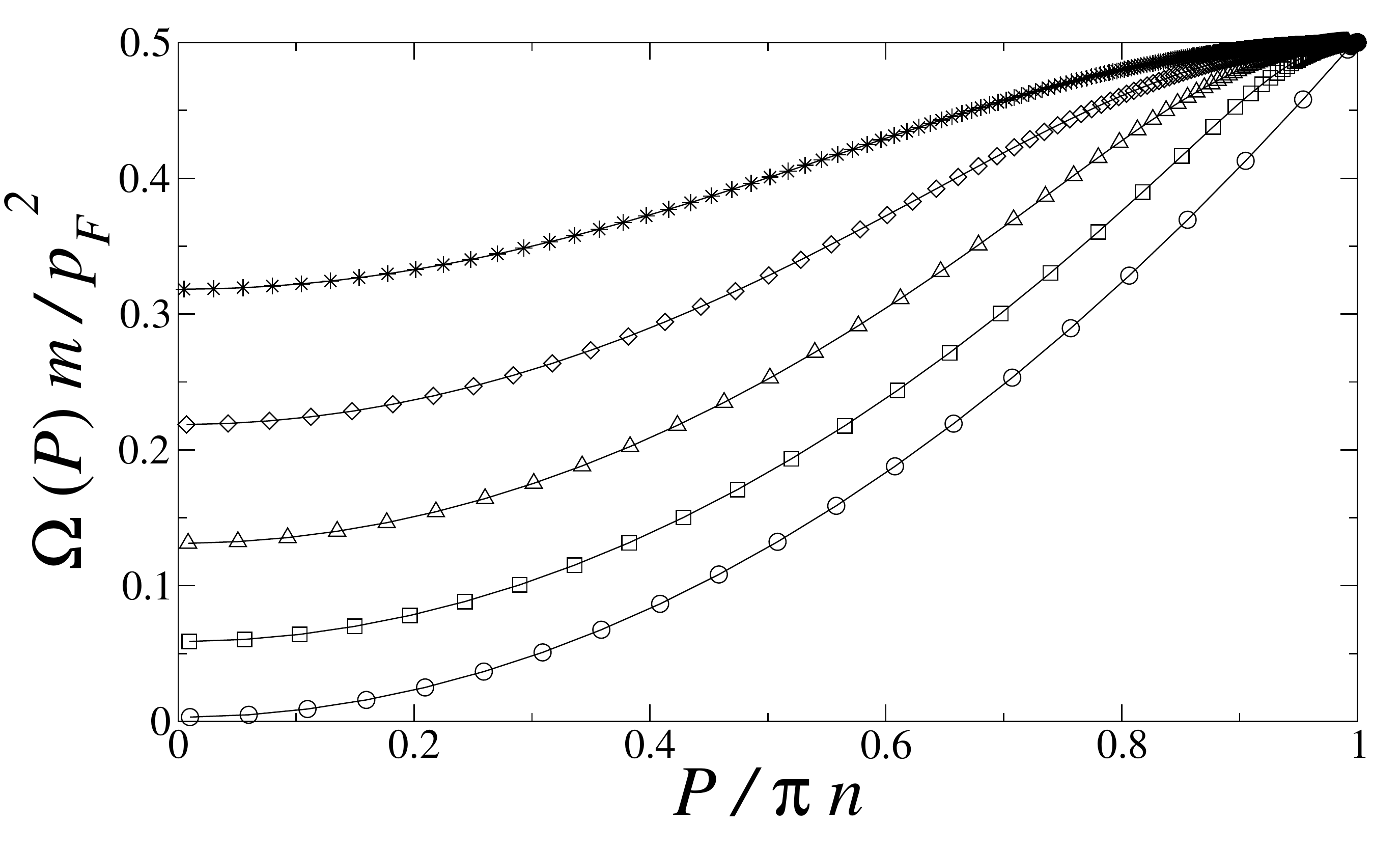}
\includegraphics[width=0.4\textwidth]{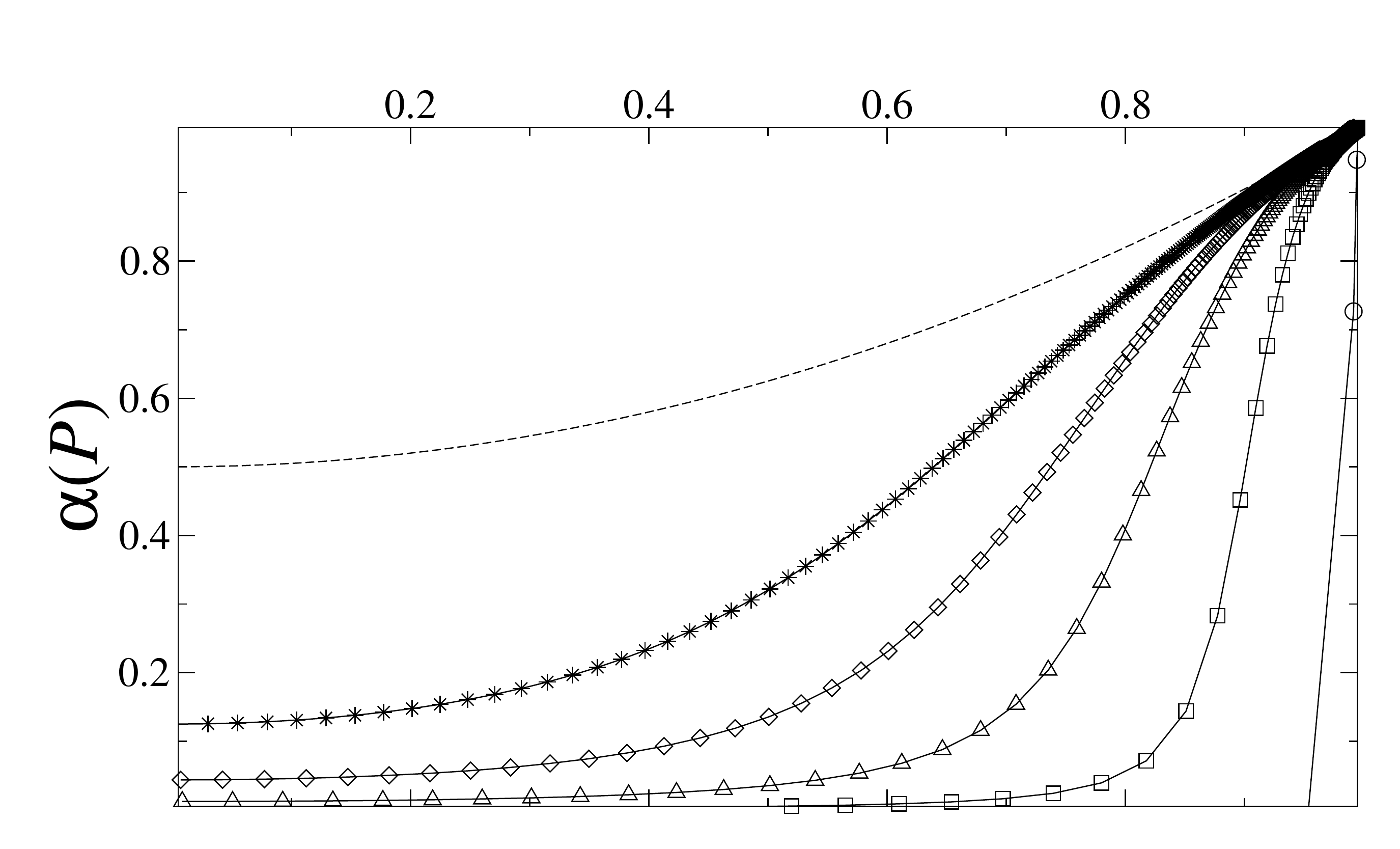}
\caption{Dispersion relation (top) and spectral function exponent (bottom) for the exactly soluble limit of an impurity of mass $M=m$ moving in a Fermi gas, corresponding to Luttinger parameter $K=1$. 
Results are given for $\gamma\equiv mu/\hbar^2 n=0.01 (\circ), 0.2(\square), 0.5 (\triangle), 1(\Diamond)$, and 2$(\star)$. The dashed curve gives the exponent $\alpha(P)=\frac{1}{2}[1+\left(\frac{P}{\pi n}\right)^2]$ valid in the $\gamma\to\infty$ limit.
 \label{fig:disp}}
\end{figure}

The extremely simple form of the eigenstates allows us to extract the behavior of the spectral function easily. Indeed it is possible to choose the $\varphi_j(x)$ so that they approach plane wave states with forward scattering phases $\delta(k)$ in the $L\to \infty$ limit~\cite{castella1993}. The form of the spectral function exponent therefore coincides with the classic result for the deep hole correlation function in the X-ray edge problem~\cite{nozieres1969}: $\alpha(P)=\left(\delta(p_F)^2+\delta(-p_F)^2\right)/\pi^2$. In the $u\to\infty$ limit one finds $\delta(\pm p_F)=\frac{\pi P}{2p_F}\mp \frac{\pi }{2}$ and thus $\alpha(P)=\frac{1}{2}[1+\left(\frac{P}{\pi n}\right)^2]$ in agreement with our general result. The form of $\Omega(P)$ and $\alpha(P)$ are shown in Fig.~\ref{fig:disp} for various coupling strengths. The persistence of edge singularities for a massive impurity in one dimension was noted some time ago for $P=0$~\cite{ogawa1992,nozieres1994}.
%
	

In summary, we have shown that the motion of an impurity in a one dimensional quantum gas has some interesting features that have their origin in singular backward scattering processes at particular momenta. This system has been recently realized in an ultracold atomic gas~\cite{Palzer2009}.

The author is extremely grateful to Fabian Essler for many useful discussions and his collaboration in the early stages of this work, as well as insightful comments from Adilet Imambekov, Alex Kamenev, and Leonid Glazman.


\end{document}